\documentclass[twocolumn,showpacs,preprintnumbers,amsmath,amssymb,prl,superscriptaddress]{revtex4}
\usepackage{epsfig}
\usepackage{graphicx}
\usepackage{dcolumn}
\usepackage{bm}
\usepackage{color} 

\begin{document}

\title{
Observation of an optical non-Fermi-liquid behavior in the heavy fermion state of YbRh$_{2}$Si$_{2}$
}

\author{S. Kimura}
 \altaffiliation[Electronic address: ]{kimura@ims.ac.jp}
\affiliation{
UVSOR Facility, Institute for Molecular Science, Okazaki 444-8585, Japan
}
\affiliation{
School of Physical Sciences, The Graduate University for Advanced Studies (SOKENDAI), Okazaki 444-8585, Japan
}
\author{J. Sichelschmidt}
\author{J. Ferstl}
\author{C. Krellner}
\author{C. Geibel}
\author{F. Steglich}
\affiliation{
Max Planck Institute for Chemical Physics of Solids, N\"othnitzer Stra\ss e 40, 01187 Dresden, Germany
}
\date{\today}

\begin{abstract} 

We report far-infrared optical properties of YbRh$_{2}$Si$_{2}$ for photon energies down to 2~meV and temperatures 0.4~--~300~K.
In the coherent heavy quasiparticle state, a {\it linear} dependence of the low-energy scattering rate on both temperature and photon energy was found.
We relate this distinct dynamical behavior different from that of Fermi liquid materials to the non-Fermi liquid nature of YbRh$_{2}$Si$_{2}$ which is due to its close vicinity to an antiferromagnetic quantum critical point.

\end{abstract}

\pacs{71.10.Hf, 71.27.+a, 78.20.-e}

\maketitle
%
%
The investigation of $4f$-containing metals by far-infrared optical spectroscopy provides valuable insight into the nature of strong electronic correlations.
This in particular holds true for heavy fermion (HF) compounds where at low temperatures a weak $4f$-conduction electron ($cf$-)hybridization generates mass-renormalized quasiparticles with a coherent ground state which is in many HF systems of the Landau Fermi liquid (LFL) type.~\cite{stewart01}
The quasiparticles influence thermodynamic quantities which are described in terms of a large effective mass $m^{*}$ exceeding the free electron mass $m_{0}$ by three orders of magnitude.
Furthermore, in typical HF materials, below a single-ion Kondo temperature ($T_{\rm K}$), the coherent state is characterized by a dynamical screening of the $4f$ magnetic moments through the conduction electrons.
Several highly correlated metals exhibit so-called non-Fermi liquid (NFL), {\it i.e.}, strong deviations from a renormalized LFL behavior when $T\rightarrow0$~K.~\cite{stewart01} The system YbRh$_{2}$Si$_{2}$ studied in this paper is one of a few clean stoichiometric HF metals with pronounced NFL behavior at ambient pressure which is related to both antiferromagnetic (AF) as well as ferromagnetic quantum critical spin fluctuations in close proximity to an AF quantum critical point (QCP).~\cite{geg02,cus03,geg05}
Those NFL effects manifest as a divergence of the $4f$-derived increment to the specific heat $\Delta C/T \propto -\ln T$ and in the electrical resistivity $\rho(T)$ showing a power law exponent close to 1 in a temperature range substantially larger than one decade and extending up to $T\simeq10$~K.~\cite{tro00}
Transport and thermodynamic properties are consistent with a single-ion Kondo temperature $T_{\rm K}=25$~K (associated with the crystalline-electric-field-derived doublet ground state~\cite{sto05}).

\indent The electrodynamical response of HF systems is characterized by an optical conductivity $\sigma(\omega)$ which follows at room temperature the {\it classical} Drude model [$\sigma(\omega)~=~N e^{2} \tau/{m^{*} (1 + \omega^{2} \tau^{2})}$; $N$: charge carrier density] with frequency independent $m^{*}$ and scattering rate $1/\tau$.~\cite{DG02}
At low temperatures, upon entering the coherent state, large deviations are observed which are caused by many-body effects.
Then a narrow, renormalized peak at zero photon energy $\hbar\omega$ = 0~eV is formed and a so-called hybridization gap appears which is related to the transition between the bonding and antibonding states resulting from the $cf$-hybridization.~\cite{deg01,mil87a,mil87b}
The coherent part of the underlying strong electron-electron correlations are treated in an {\it extended} Drude model by renormalized and frequency dependent $m^{*}(\omega)/m_{0}$ and $1/\tau(\omega)$;~\cite{web86,awa93,kim94,deg99}
\[
\frac{m^{*}(\omega)}{m_{0}} = \frac{N e^2}{m_0 \omega} \cdot Im\left(\frac{1}{\tilde{\sigma}(\omega)}\right), 
\frac{1}{\tau(\omega)} = \frac{N e^2}{m_0} \cdot Re\left(\frac{1}{\tilde{\sigma}(\omega)}\right).
\]
Here, $\tilde{\sigma}(\omega)$ is the complex optical conductivity derived from the Kramers-Kronig analysis (KKA) of the reflectivity spectrum $R(\omega)$.
The LFL theory predicts a dynamical scattering rate $1/\tau(\omega)~\propto~(2\pi k_{\rm B}T)^2+(\hbar\omega)^2$ which also accounts for the electrical resistivity, $\rho(T)$, growing quadratically with temperature~\cite{deg99}.
The $(\hbar\omega)^2$ behavior is indicated in $1/\tau(\omega)$ of many renormalized LFL metals, e.g. YbAl$_3$~\cite{oka04}, CePd$_3$~\cite{web86}, and CeAl$_3$~\cite{awa93}.
At the same time, $m^*(\omega)$ increases with decreasing $T$ and $\omega$ indicating the formation of heavy quasiparticles at low temperatures. NFL behavior in optical properties is typically indicated by a linear frequency dependence of $1/\tau(\omega)$.~\cite{deg99}
Up to now, optical NFL effects were explicitly investigated for correlated materials whose NFL state is believed to be related to disorder (several U-based Kondo alloys) or to two-channel Kondo physics (UBe$_{13}$).~\cite{deg99}

\indent Yet, to our knowledge, the optical properties of a heavy-fermion NFL state due to spin fluctuations in close proximity to a QCP, as is the case for YbRh$_{2}$Si$_{2}$, have not been investigated so far.
As shown by our preliminary optical experiments on YbRh$_{2}$Si$_{2}$ the $T$-linear NFL behavior of the zero-frequency resistivity, $\rho_{\rm DC}(T)$, is also reflected in $\sigma(\omega,T)$ for $T<20$~K, $\hbar\omega<20$~meV and $\omega\tau\gg1$, assuming a frequency independent $\tau$ consistent with a {\it classical} Drude approximation of the data.~\cite{kim04}
This behavior was interpreted as the temperature dependence of a renormalized scattering rate of a Drude peak whose tail at $T=2.7$~K was observable just above the lowest measured energy of 10~meV.
Moreover, a peak at around 0.2~eV, visible already at $T=300$~K and gradually developing with decreasing temperature, appears beyond a pseudogap-like structure similar to that reported for several other Kondo-lattice systems.~\cite{deg01,oka04,web86,men05}

\indent Here we report the extension of our optical investigations down to energies of 2~meV and temperatures down to 0.4~K.
This allowed us to obtain yet inaccessible information on the low-energy HF optical response of YbRh$_{2}$Si$_{2}$ and provides a detailed characterization of the electrodynamic NFL properties.
In particular the low-energy heavy quasiparticle excitations could be analyzed within the {\it extended} Drude model which yields $m^*(\omega,T)$ and $1/\tau(\omega,T)$.

%
%
\indent Near-normal incident $R(\omega)$ spectra were acquired in a very wide photon-energy region of 2~meV -- 30~eV to ensure an accurate KKA.
We investigated the tetragonal $ab$-plane of two single crystalline samples with as-grown sample surfaces and sizes of $2.2 \times 1.5 \times 0.1$ mm$^3$ and $3.5 \times 4.2 \times 0.5$ mm$^3$, respectively.
The preparation as well as the magnetic and transport properties has been described elsewhere.~\cite{tro00,geg02,cus03}
The high quality of the single crystals is evidenced by a residual resistivity ratio of $\rho_{\rm 300K}/\rho_0\simeq 65$ $(\rho_0\simeq1\mu\Omega {\rm cm})$ and a very sharp anomaly in the specific heat at $T = T_{\rm N}$.~\cite{cus03}
Rapid-scan Fourier spectrometers of Martin-Puplett and Michelson type were used at photon energies of 2--30~meV and 0.01--1.5~eV, respectively, at sample temperatures between 0.4--300~K using a $^4$He ($T\rightarrow 5.5$~K) and a $^3$He ($T\rightarrow 0.4$~K) cryostat.
To obtain $R(\omega)$, a reference spectrum was measured by using the sample surface evaporated {\it in-situ} with gold.
At $T=300$~K, $R(\omega)$ was measured for energies 1.2--30~eV by using synchrotron radiation.~\cite{fuk01}
In order to obtain $\sigma(\omega)$ via a KKA of $R(\omega)$ the spectra were extrapolated below 2~meV with $R(\omega)=1-(2\omega/\pi \sigma_{DC})^{1/2}$ and above 30~eV with a free-electron approximation $R(\omega) \propto \omega^{-4}$.~\cite{DG02}
 
\begin{figure}[t]
\begin{center}
\includegraphics[width=0.35\textwidth]{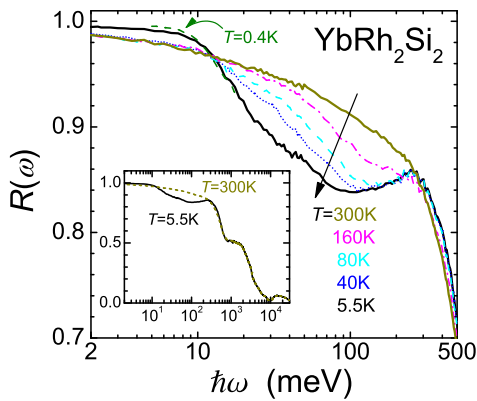}
\end{center}
\caption{
(Color online) Temperature dependence of the reflectivity spectrum $R(\omega)$ in the photon energy range of 2 -- 500~meV.
Inset: $R(\omega)$ at 5.5 and 300~K in the complete accessible range of photon energies up to 30~eV.
}
\label{fig1}
\end{figure}
\indent The temperature dependence of the $R(\omega)$ spectra of YbRh$_{2}$Si$_{2}$ is shown in Fig.~\ref{fig1}.
The inset shows an extended energy region where above 500~meV $R(\omega)$ is dominated by interband transitions.
In this study, we focus only on the intraband transition region below 500~meV where the spectra display a strong temperature dependence.
With decreasing temperature $R(\omega)$ gets strongly suppressed, creating a dip structure at around 100~meV.
Simultaneously, below 12~meV, $R(\omega)$ approaches unity with decreasing temperature.
These pronounced temperature dependences are typical for HF compounds.~\cite{deg99}
Most clear coincidence is found when comparing the optical properties of YbRh$_{2}$Si$_{2}$ with those of the intermediate-valent compound YbAl$_{3}$.~\cite{oka04}
Their low-temperature, low-energy shapes of $R(\omega)$ are very similar, albeit a weaker temperature dependence is found for YbAl$_{3}$ reflecting its much stronger $cf$-hybridization which underlies its intermediate-valence behavior.
However, very similar to $R(\omega)$ of YbRh$_{2}$Si$_{2}$ at $T=300$~K, the $R(\omega)$ of the non-magnetic reference compound LaAl$_{3}$ does not show any dip-structure.
Therefore, as already identified for YbAl$_{3}$, the pronounced low-temperature dip in $R(\omega)$ of YbRh$_{2}$Si$_{2}$ can be related to Yb-$4f$ electronic states near the Fermi energy.
By decreasing the temperature, due to $cf$-hybridization, the character of the 4f states changes from localized to itinerant where optical transitions between the $cf$-hybridization states are expected.~\cite{deg01}
This is consistent with the observed $R(\omega)$ dip structure and its temperature evolution in YbRh$_{2}$Si$_{2}$.

\begin{figure}[t]
\begin{center}
\includegraphics[width=0.35\textwidth]{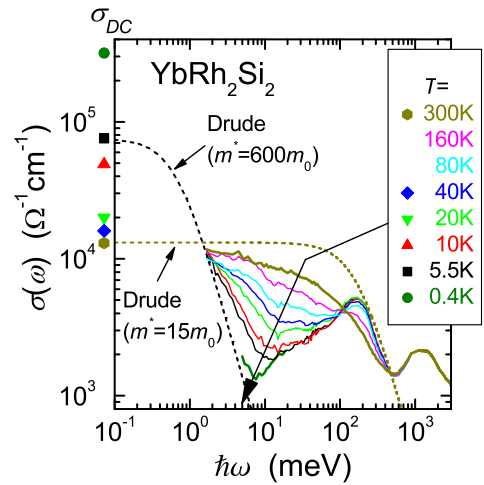}
\end{center}
\caption{
(Color online) Temperature dependence of the optical conductivity $\sigma(\omega)$ (solid lines) with corresponding direct current conductivity ($\sigma_{DC}$, symbols).
Dashed lines: {\it Classical} Drude model with implicit Drude masses $m^*$ as indicated.
Corresponding $\sigma_{DC}$ and carrier densities (derived from the Hall coefficient) were used.
}
\label{fig2}
\end{figure}
\begin{figure}[t]
\begin{center}
\includegraphics[width=0.3\textwidth]{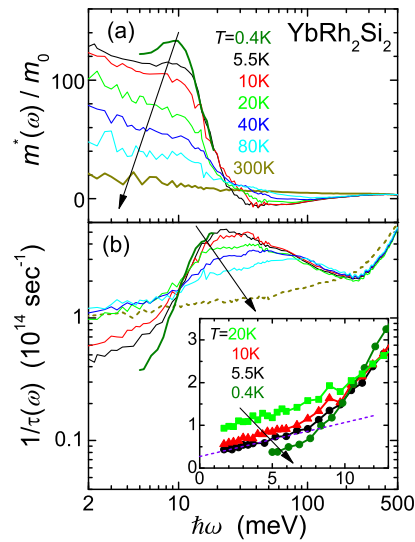}
\end{center}
\caption{
(Color online) Temperature dependence of (a) the effective mass relative to the free electron mass, $m^{*}(\omega)/m_{0}$, and (b) the scattering rate $1/\tau(\omega)$ as a function of photon energy $\hbar\omega$.
Inset of (b) is the low energy part of $1/\tau(\omega)$.
Dashed line emphasizes a $1/\tau\propto\hbar\omega$ behavior.
}
\label{fig3}
\end{figure}

\indent The KKA of $R(\omega)$ yields optical quantities as shown in Fig.~\ref{fig2}.
At $T=300$~K, $\sigma(\omega)$ shows normal metallic behavior, {\it i.e.}, a monotonic decrease with increasing photon energy, and a zero-energy extrapolation consistent with $\sigma_{DC}$ (symbols at left axis of Fig.~\ref{fig2}).
However, as shown by the dashed line in Fig.~\ref{fig2}, the experimental $\sigma(\omega)$ is poorly represented by a {\it classical} Drude fit [with parameters $m^{*} = 15 m_{0} $, $1/\tau = 4.0 \cdot 10^{13}$ sec$^{-1}$, $N = 2.7 \cdot 10^{22}$~cm$^{-3}$ (Hall effect result~\cite{pas04})].
This discrepancy indicates that the scattering rate depends on photon energy, as discussed below and as shown in Fig.~\ref{fig3}b. With decreasing temperature, a pseudogap-like suppression of $\sigma(\omega)$ appears below 100~meV with a simultaneous increase in $\sigma_{DC}$. A minimum of $\sigma(\omega)$ develops whose position continuously decreases towards low temperatures.
The onset temperature of pseudogap formation between 80 and 160~K corresponds to the maximum of $\rho_{\rm DC}(T)$ at $T_{\rm coh}=120$~K which marks the onset of coherence effects upon $cf$-hybridization.
This suggests that the temperature dependence of $\sigma(\omega)$ is indeed related to the formation of heavy quasiparticles and the formation of a minimum in $\sigma(\omega)$ may be associated with a heavy plasma mode.

\indent As already indicated from the above discussion, the energy and temperature behavior of the optical conductivity implies that highly energy dependent $m^{*}$ and $1/\tau(\omega)$ are involved.
For example, $\sigma(\omega)$ at $T=5.5$~K cannot be represented by energy-independent values of both $m^{*}$ and $1/\tau$ within a {\it classical} Drude curve as shown in Fig.~\ref{fig2}.
Moreover, due to the different temperature dependences in $\sigma(\omega)$ and $\sigma_{DC}$, the {\it classical} Drude analysis emphasizes the need for strongly temperature dependent and, at low temperatures, very heavy effective masses ($m_{\rm Drude}^{*} = 600~m_{0}$, $1/\tau = 1.6 \cdot 10^{11}$ sec$^{-1}$ at 5.5~K).
In general, such behavior of the optical mass and scattering rate reflects electron-electron scattering or electron scattering off spin fluctuations.
In case of HF compounds, a many-body effect due to the $cf$-hybridization is effective at low energies and temperatures where the conduction electrons are scattered resonantly off the hybridized charge carriers.~\cite{deg01}

%
\indent Such scattering process is reflected in the temperature- and photon-energy dependences of $m^{*}$ and $1/\tau$ which we obtained from an {\it extended} Drude analysis and which are shown in Fig.~\ref{fig3} for energies lower than the interband transition spectrum.
At $T=300$~K both $m^{*}(\omega)/m_{0}$ and $1/\tau(\omega)$ are almost constant, with values of about $15m_{0}$ and $1\cdot10^{14}$ sec$^{-1}$, respectively.
Therefore, it is not surprising that $\sigma(\omega)$ at 300~K clearly contains the features of a {\it classical} Drude model as shown in Fig.~\ref{fig2}.
With decreasing temperature from 300~K to 0.4~K, $m^{*}(\omega)/m_{0}$ below $\simeq20$~meV monotonically increases and exceeds values of 130.
Clearly, this enhancement can be related to the HF state formation in YbRh$_{2}$Si$_{2}$ as the enhancement of $m^*(\omega)$ occurs at energies comparable to $k_{\rm B}T_{\rm coh}$. Interesting to note, below 10~meV, $m^{*}(\omega)/m_{0}$ does not seem to saturate with decreasing temperature and energy but rather increases continuously.
We speculate that this behavior indicates an energy equivalence to the electron effective mass temperature divergence to infinity as observed in the electronic specific heat.~\cite{tro00,geg02,cus03}
The appearance of a negative mass at energies above $\simeq30$~meV and at low $T$ is caused by a positive $\varepsilon_{1}(\omega)$ indicating a heavy plasma mode (not shown).
Equivalently, one may relate transitions across the hybridization gap to the observed negative optical mass.
Such behavior is observed in many other heavy-fermion materials.~\cite{bon88, men05, dre02, dor01}

The $m^{*}(\omega)/m_{0}$ enhancement with decreasing temperature is accompanied by a formation of a broad peak in $1/\tau(\omega)$ in the energy region where the pseudogap-like suppression of $\sigma(\omega)$ appears as shown in Fig.~\ref{fig3}b.
It is related to the process of mass renormalization as, at 0.4~K, the $1/\tau(\omega)$ reaches the maximum position of $\simeq22$~meV which corresponds to the onset of the $m^{*}(\omega)/m_{0}$ enhancement.
Again, transitions across the hybridization gap lead to such enhanced dynamical scattering rates reflecting the particular quasiparticle excitation in accord with hybridization-gap scenarios for HF-derived optical properties.~\cite{mil87a,deg01,awa93}
As shown in the inset of Fig.~\ref{fig3}b the HF state is characterized by $1/\tau~\propto~\hbar\omega$ for energies up to $\simeq7$~meV which is a pronounced NFL behavior, see the dashed line for the data at 5.5~K. 
It is worth to remind that in stoichiometric YbRh$_{2}$Si$_{2}$ NFL effects due to disorder can be excluded.~\cite{tro00}
Therefore, we attribute the low-energy linear in $\omega$ behavior of $1/\tau(\omega)$ to spin fluctuations due to the close vicinity to the QCP. 

\begin{figure}[t]
\begin{center}
\includegraphics[width=0.3\textwidth]{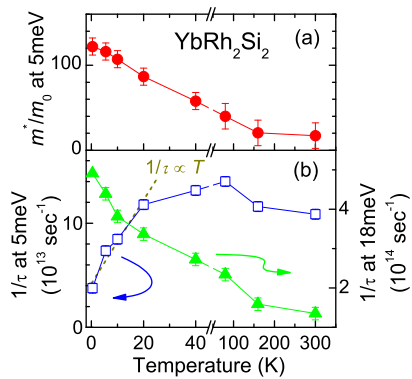}
\end{center}
\caption{
(Color online) Temperature dependence of (a) the dynamical mass $m^{*}(T)/m_{0}$ and (b) the scattering rate $1/\tau(\omega,T)$ at specific photon energies as indicated.
Dashed line in (b) shows a non-Fermi liquid $1/\tau\propto T$ behavior.
}
\label{fig4}
\end{figure}
\indent The extended Drude description of the optical properties of correlated electron systems yields the energy dependence of the renormalization effects.
In the low-energy limit the frequency dependence of both $m^*$ and $1/\tau$ should resemble their temperature dependence.~\cite{dre02}
This expectation is satisfied when comparing the data of Fig.~\ref{fig3} with Fig.~\ref{fig4}.
The latter shows the temperature dependence of $m^{*}(T)/m_{0}$ at 5~meV and that of $1/\tau(T)$ at 5~meV and at 18~meV obtained from Fig.~\ref{fig3}.
Note that the $m^{*}(T)/m_{0}$ enhancement occurs below 160~K which roughly corresponds to the onset energy of mass enhancement.
Similar to $m^{*}(\omega)/m_{0}$, $m^{*}(T)/m_{0}$ does not saturate even at the lowest accessible temperature below $T_{\rm K}=25$~K.
However, in contrast to the divergence of the electronic specific heat coefficient $\Delta C/T \propto -\ln T$, $m^{*}(T,\omega)/m_{0}$ shows an almost linear increase towards low temperatures, at least for photon energies down to 5~meV.
From this discrepancy, we anticipate that a divergence of the optical mass renormalization may occur below the single-ion Kondo energy scale of $k_{\rm B}T_{\rm K}=2$~meV.
The mass enhancement with decreasing temperature corresponds to a continuous increase of $1/\tau(T)$ at 18~meV as shown in Fig.~\ref{fig4}b.
However, below $T_{\rm K}$ the increase of $1/\tau(T)$ at 18~meV becomes stronger.
At the same time, $1/\tau(T)$ at 5~meV assumes a NFL temperature dependence which is approximately linear as the dashed line emphasizes in Fig.~\ref{fig4}b.
Therefore, at the single-ion Kondo temperature $T_{\rm K}$, the charge dynamics changes while a single ion Kondo scenario fails to explain the magnetic properties of YbRh$_{2}$Si$_{2}$ below $T_{\rm K}$ (large fluctuating $4f$-magnetic moments~\cite{cus03} and a sharp electron spin resonance line~\cite{sic03}).

\indent At temperatures near 80~K, $m^{*}(T)/m_{0}$ starts to get enhanced and $1/\tau(T)$ shows a kink or a small peak.
This temperature range corresponds to that at which both the $^{29}$Si-NMR Knight shift and relaxation rate show an anomaly,~\cite{ish02} indicating a change in the magnetic characteristics at 80~K.
For a proper interpretation of the peak in $1/\tau(T)$, carrier scattering by phonons should also be taken into account.

%
%
\indent In conclusion, we found distinct electrodynamical non-Fermi liquid behavior of the low-energy charge dynamics of clean ({\it i.e.},atomically ordered, stoichiometric) YbRh$_{2}$Si$_{2}$.
We relate our results to the close proximity of YbRh$_{2}$Si$_{2}$ to an antiferromagnetic quantum critical point as the latter is the origin of the pronounced NFL effects of thermodynamic and transport properties.~\cite{tro00,geg02,cus03}
Our findings were  accomplished by measuring the temperature dependence of the optical conductivity of YbRh$_{2}$Si$_{2}$ down to $T=0.4$~K in the photon energy range 2~meV -- 30~eV.
From an extended Drude analysis, the scattering rate below $\hbar\omega\simeq7$~meV and below $T\simeq20$~K is consistent with a NFL linear proportionality both to the photon energy and temperature.
Moreover, towards low temperatures, clear signatures of heavy fermion behavior are found: formation of an interband peak at 0.2~eV and a heavy plasmon mode below 30~meV which both can be related to $cf$-hybridization.
The low-temperature optical effective mass is strongly enhanced below 20~meV and continues to increase down to the accessible lowest energies (2~meV) and temperatures (0.4~K).

\indent We would like to thank Q. Si and O. Sakai for fruitful discussions.
This work was a joint studies program of the Institute for Molecular Science and was partially supported by Grants-in-Aid for Scientific Research (Grant No.~18340110) from MEXT of Japan and by DFG under the auspices of SFB 463 of Germany.
%
%

\begin{thebibliography}{21}
\expandafter\ifx\csname natexlab\endcsname\relax\def\natexlab#1{#1}\fi
\expandafter\ifx\csname bibnamefont\endcsname\relax
  \def\bibnamefont#1{#1}\fi
\expandafter\ifx\csname bibfnamefont\endcsname\relax
  \def\bibfnamefont#1{#1}\fi
\expandafter\ifx\csname citenamefont\endcsname\relax
  \def\citenamefont#1{#1}\fi
\expandafter\ifx\csname url\endcsname\relax
  \def\url#1{\texttt{#1}}\fi
\expandafter\ifx\csname urlprefix\endcsname\relax\def\urlprefix{URL }\fi
\providecommand{\bibinfo}[2]{#2}
\providecommand{\eprint}[2][]{\url{#2}}

\bibitem[{\citenamefont{Stewart}(2001)}]{stewart01}
\bibinfo{author}{\bibfnamefont{G.~R.} \bibnamefont{Stewart}},
  \bibinfo{journal}{Rev.\ Mod.\ Phys.} \textbf{\bibinfo{volume}{73}},
  \bibinfo{pages}{797} (\bibinfo{year}{2001}).

\bibitem[{\citenamefont{Gegenwart et~al.}(2002)\citenamefont{Gegenwart,
 Custers, Geibel, Neumaier, Tayama, Tenya, Trovarelli, and Steglich}}]{geg02}
\bibinfo{author}{\bibfnamefont{P.}~\bibnamefont{Gegenwart}},
\bibinfo{author}{\bibfnamefont{J.}~\bibnamefont{Custers}},
\bibinfo{author}{\bibfnamefont{C.}~\bibnamefont{Geibel}},
\bibinfo{author}{\bibfnamefont{K.}~\bibnamefont{Neumaier}},
\bibinfo{author}{\bibfnamefont{T.}~\bibnamefont{Tayama}},
\bibinfo{author}{\bibfnamefont{K.}~\bibnamefont{Tenya}},
\bibinfo{author}{\bibfnamefont{O.}~\bibnamefont{Trovarelli}},
\bibnamefont{and}
\bibinfo{author}{\bibfnamefont{F.}~\bibnamefont{Steglich}},
  \bibinfo{journal}{Phys.\ Rev.\ Lett.} \textbf{\bibinfo{volume}{89}},
  \bibinfo{pages}{056402} (\bibinfo{year}{2002}).

\bibitem[{\citenamefont{Custers et~al.}(2003)\citenamefont{Custers, Gegenwart,
 Wilhelm, Neumaier, Tokiwa, Trovarelli, Geibel, Steglich, Pepin, and Coleman}}]{cus03}
\bibinfo{author}{\bibfnamefont{J.}~\bibnamefont{Custers}},
\bibinfo{author}{\bibfnamefont{P.}~\bibnamefont{Gegenwart}},
\bibinfo{author}{\bibfnamefont{H.}~\bibnamefont{Wilhelm}},
\bibinfo{author}{\bibfnamefont{K.}~\bibnamefont{Neumaier}},
\bibinfo{author}{\bibfnamefont{Y.}~\bibnamefont{Tokiwa}},
\bibinfo{author}{\bibfnamefont{O.}~\bibnamefont{Trovarelli}},
\bibinfo{author}{\bibfnamefont{C.}~\bibnamefont{Geibel}},
\bibinfo{author}{\bibfnamefont{F.}~\bibnamefont{Steglich}},
\bibinfo{author}{\bibfnamefont{C.}~\bibnamefont{P$\acute{e}$pin}},
\bibnamefont{and}
\bibinfo{author}{\bibfnamefont{P.}~\bibnamefont{Coleman}},
  \bibinfo{journal}{Nature} \textbf{\bibinfo{volume}{424}},
  \bibinfo{pages}{524} (\bibinfo{year}{2003}).

\bibitem[{\citenamefont{Gegenwart et~al.}(2005)\citenamefont{Gegenwart,
 Custers, Tokiwa, Geibel, and Steglich}}]{geg05}
\bibinfo{author}{\bibfnamefont{P.}~\bibnamefont{Gegenwart}},
\bibinfo{author}{\bibfnamefont{J.}~\bibnamefont{Custers}},
\bibinfo{author}{\bibfnamefont{Y.}~\bibnamefont{Tokiwa}},
\bibinfo{author}{\bibfnamefont{C.}~\bibnamefont{Geibel}},
\bibnamefont{and}
\bibinfo{author}{\bibfnamefont{F.}~\bibnamefont{Steglich}},
  \bibinfo{journal}{Phys.\ Rev.\ Lett.} \textbf{\bibinfo{volume}{94}},
  \bibinfo{pages}{076402} (\bibinfo{year}{2005}).

\bibitem[{\citenamefont{Trovarelli et~al.}(2000)\citenamefont{Trovarelli,
 Geibel, Mederle, Langhammer, Grosche, Gegenwart, Lang, Sparn, and
 Steglich}}]{tro00}
\bibinfo{author}{\bibfnamefont{O.}~\bibnamefont{Trovarelli}},
\bibinfo{author}{\bibfnamefont{C.}~\bibnamefont{Geibel}},
\bibinfo{author}{\bibfnamefont{S.}~\bibnamefont{Mederle}},
\bibinfo{author}{\bibfnamefont{C.}~\bibnamefont{Langhammer}},
\bibinfo{author}{\bibfnamefont{F.~M.} \bibnamefont{Grosche}},
\bibinfo{author}{\bibfnamefont{P.}~\bibnamefont{Gegenwart}},
\bibinfo{author}{\bibfnamefont{M.}~\bibnamefont{Lang}},
\bibinfo{author}{\bibfnamefont{G.}~\bibnamefont{Sparn}},
\bibnamefont{and}
\bibinfo{author}{\bibfnamefont{F.}~\bibnamefont{Steglich}},
  \bibinfo{journal}{Phys.\ Rev.\ Lett.} \textbf{\bibinfo{volume}{85}},
  \bibinfo{pages}{626} (\bibinfo{year}{2000}).


\bibitem[{\citenamefont{Stockert et~al.}(2006)\citenamefont{Stockert, Koza,
 Ferstl, Murani, Geibel, and Steglich}}]{sto05}
\bibinfo{author}{\bibfnamefont{O.}~\bibnamefont{Stockert}},
\bibinfo{author}{\bibfnamefont{M.~M.} \bibnamefont{Koza}},
\bibinfo{author}{\bibfnamefont{J.}~\bibnamefont{Ferstl}},
\bibinfo{author}{\bibfnamefont{A.~P.} \bibnamefont{Murani}},
\bibinfo{author}{\bibfnamefont{C.}~\bibnamefont{Geibel}},
\bibnamefont{and}
\bibinfo{author}{\bibfnamefont{F.}~\bibnamefont{Steglich}},
 \bibinfo{journal}{Physica B} \textbf{\bibinfo{volume}{378-380}},
 \bibinfo{pages}{157} (\bibinfo{year}{2006}).
  

\bibitem[{\citenamefont{Dressel and Gr\"uner}(2002)}]{DG02}
\bibinfo{author}{\bibfnamefont{M.}~\bibnamefont{Dressel}}
\bibnamefont{and}
\bibinfo{author}{\bibfnamefont{G.}~\bibnamefont{Gr\"uner}},
  \emph{\bibinfo{title}{Electrodynamics of Solids}}
  (\bibinfo{publisher}{Cambridge Univ. Press}, \bibinfo{address}{Cambridge},
  \bibinfo{year}{2002}).

\bibitem[{\citenamefont{Degiorgi et~al.}(2001)\citenamefont{Degiorgi, Anders,
 and Gr\"uner}}]{deg01}
\bibinfo{author}{\bibfnamefont{L.}~\bibnamefont{Degiorgi}},
\bibinfo{author}{\bibfnamefont{F.}~\bibnamefont{Anders}},
\bibnamefont{and}
\bibinfo{author}{\bibfnamefont{G.}~\bibnamefont{Gr\"uner}},
  \bibinfo{journal}{Eur.\ Phys.\ J.\ B} \textbf{\bibinfo{volume}{19}},
  \bibinfo{pages}{167} (\bibinfo{year}{2001}).

\bibitem[{\citenamefont{Millis and Lee}(1987)}]{mil87a}
\bibinfo{author}{\bibfnamefont{A.~J.} \bibnamefont{Millis}}
\bibnamefont{and}
\bibinfo{author}{\bibfnamefont{P.~A.} \bibnamefont{Lee}},
  \bibinfo{journal}{Phys.\ Rev.\ B} \textbf{\bibinfo{volume}{35}},
  \bibinfo{pages}{3394} (\bibinfo{year}{1987}).

\bibitem[{\citenamefont{Millis et~al.}(1987)\citenamefont{Millis, Lavagna, and
 Lee}}]{mil87b}
\bibinfo{author}{\bibfnamefont{A.~J.} \bibnamefont{Millis}},
\bibinfo{author}{\bibfnamefont{M.}~\bibnamefont{Lavagna}},
\bibnamefont{and}
\bibinfo{author}{\bibfnamefont{P.~A.} \bibnamefont{Lee}},
  \bibinfo{journal}{Phys.\ Rev.\ B} \textbf{\bibinfo{volume}{36}},
  \bibinfo{pages}{864} (\bibinfo{year}{1987}).

\bibitem[{\citenamefont{Degiorgi}(1999)}]{deg99}
\bibinfo{author}{\bibfnamefont{L.}~\bibnamefont{Degiorgi}},
  \bibinfo{journal}{Rev.\ Mod.\ Phys.} \textbf{\bibinfo{volume}{71}},
  \bibinfo{pages}{687} (\bibinfo{year}{1999}).

\bibitem[{\citenamefont{Webb et~al.}(1986)\citenamefont{Webb, Sievers, and
 Mihalisin}}]{web86}
\bibinfo{author}{\bibfnamefont{B.~C.} \bibnamefont{Webb}},
\bibinfo{author}{\bibfnamefont{A.~J.} \bibnamefont{Sievers}},
\bibnamefont{and}
\bibinfo{author}{\bibfnamefont{T.}~\bibnamefont{Mihalisin}},
  \bibinfo{journal}{Phys.\ Rev.\ Lett.} \textbf{\bibinfo{volume}{57}},
  \bibinfo{pages}{1951} (\bibinfo{year}{1986}).

\bibitem[{\citenamefont{Awasthi et~al.}(1993)\citenamefont{Awasthi, Degiorgi,
 {G. Gr\"uner}, Dalichaouch, and Maple}}]{awa93}
\bibinfo{author}{\bibfnamefont{A.~M.} \bibnamefont{Awasthi}},
\bibinfo{author}{\bibfnamefont{L.}~\bibnamefont{Degiorgi}},
\bibinfo{author}{\bibnamefont{{G. Gr\"uner}}},
\bibinfo{author}{\bibfnamefont{Y.}~\bibnamefont{Dalichaouch}},
\bibnamefont{and}
\bibinfo{author}{\bibfnamefont{M.~B.} \bibnamefont{Maple}},
  \bibinfo{journal}{Phys.\ Rev.\ B} \textbf{\bibinfo{volume}{48}},
  \bibinfo{pages}{10692} (\bibinfo{year}{1993}).

\bibitem[{\citenamefont{Kimura et~al.}(1994)\citenamefont{Kimura, Nanba, Kunii,
 and Kasuya}}]{kim94}
\bibinfo{author}{\bibfnamefont{S.}~\bibnamefont{Kimura}},
\bibinfo{author}{\bibfnamefont{T.}~\bibnamefont{Nanba}},
\bibinfo{author}{\bibfnamefont{S.}~\bibnamefont{Kunii}},
\bibnamefont{and}
\bibinfo{author}{\bibfnamefont{T.}~\bibnamefont{Kasuya}},
  \bibinfo{journal}{Phys.\ Rev.\ B} \textbf{\bibinfo{volume}{50}},
  \bibinfo{pages}{1406} (\bibinfo{year}{1994}).

\bibitem[{\citenamefont{Okamura et~al.}(2004)\citenamefont{Okamura, Michizawa,
 Nanba, and Ebihara}}]{oka04}
\bibinfo{author}{\bibfnamefont{H.}~\bibnamefont{Okamura}},
\bibinfo{author}{\bibfnamefont{T.}~\bibnamefont{Michizawa}},
\bibinfo{author}{\bibfnamefont{T.}~\bibnamefont{Nanba}},
\bibnamefont{and}
\bibinfo{author}{\bibfnamefont{T.}~\bibnamefont{Ebihara}},
  \bibinfo{journal}{J.\ Phys.\ Soc.\ Jpn.} \textbf{\bibinfo{volume}{73}},
  \bibinfo{pages}{2045} (\bibinfo{year}{2004}).

\bibitem[{\citenamefont{Kimura et~al.}(2004)\citenamefont{Kimura, Nishi,
  Sichelschmidt, Voevodin, Ferstl, Geibel, and Steglich}}]{kim04}
\bibinfo{author}{\bibfnamefont{S.}~\bibnamefont{Kimura}},
\bibinfo{author}{\bibfnamefont{T.}~\bibnamefont{Nishi}},
\bibinfo{author}{\bibfnamefont{J.}~\bibnamefont{Sichelschmidt}},
\bibinfo{author}{\bibfnamefont{V.}~\bibnamefont{Voevodin}},
\bibinfo{author}{\bibfnamefont{J.}~\bibnamefont{Ferstl}},
\bibinfo{author}{\bibfnamefont{C.}~\bibnamefont{Geibel}},
\bibnamefont{and}
\bibinfo{author}{\bibfnamefont{F.}~\bibnamefont{Steglich}},
  \bibinfo{journal}{J. Magn. Magn. Mat.} \textbf{\bibinfo{volume}{272-276}},
  \bibinfo{pages}{36} (\bibinfo{year}{2004}).

\bibitem[{\citenamefont{Mena et~al.}(2005)\citenamefont{Mena, {D.\ van\ der\ Marel},
 and Sarrao}}]{men05}
\bibinfo{author}{\bibfnamefont{F.~P.} \bibnamefont{Mena}},
\bibinfo{author}{\bibnamefont{{D.\ van\ der\ Marel}}},
\bibnamefont{and}
\bibinfo{author}{\bibfnamefont{J.~L.} \bibnamefont{Sarrao}},
  \bibinfo{journal}{Phys.\ Rev.\ B} \textbf{\bibinfo{volume}{72}},
  \bibinfo{pages}{045119} (\bibinfo{year}{2005}).

\bibitem[{\citenamefont{Fukui et~al.}(2001)\citenamefont{Fukui, Miura,
 Nakagawa, Shimoyama, Nakagawa, Okamura, Nanba, Hasumoto,
 and Kinoshita}}]{fuk01}
\bibinfo{author}{\bibfnamefont{K.}~\bibnamefont{Fukui}},
\bibinfo{author}{\bibfnamefont{H.}~\bibnamefont{Miura}},
\bibinfo{author}{\bibfnamefont{H.}~\bibnamefont{Nakagawa}},
\bibinfo{author}{\bibfnamefont{I.}~\bibnamefont{Shimoyama}},
\bibinfo{author}{\bibfnamefont{K.}~\bibnamefont{Nakagawa}},
\bibinfo{author}{\bibfnamefont{H.}~\bibnamefont{Okamura}},
\bibinfo{author}{\bibfnamefont{T.}~\bibnamefont{Nanba}},
\bibinfo{author}{\bibfnamefont{M.}~\bibnamefont{Hasumoto}},
\bibnamefont{and}
\bibinfo{author}{\bibfnamefont{T.}~\bibnamefont{Kinoshita}},
  \bibinfo{journal}{Nucl. Instrum. Methods Phys. Res. A}
  \textbf{\bibinfo{volume}{467-468}},
  \bibinfo{pages}{601} (\bibinfo{year}{2001}).

\bibitem[{\citenamefont{Paschen et~al.}(2004)\citenamefont{Paschen,
 {T. L\"uhmann}, Wirth, Gegenwart, Trovarelli, Geibel, Steglich,
 Coleman, and Si}}]{pas04}
\bibinfo{author}{\bibfnamefont{S.}~\bibnamefont{Paschen}},
\bibinfo{author}{\bibnamefont{{T. L\"uhmann}}},
\bibinfo{author}{\bibfnamefont{S.}~\bibnamefont{Wirth}},
\bibinfo{author}{\bibfnamefont{P.}~\bibnamefont{Gegenwart}},
\bibinfo{author}{\bibfnamefont{O.}~\bibnamefont{Trovarelli}},
\bibinfo{author}{\bibfnamefont{C.}~\bibnamefont{Geibel}},
\bibinfo{author}{\bibfnamefont{F.}~\bibnamefont{Steglich}},
\bibinfo{author}{\bibfnamefont{P.}~\bibnamefont{Coleman}},
\bibnamefont{and}
\bibinfo{author}{\bibfnamefont{Q.}~\bibnamefont{Si}},
  \bibinfo{journal}{Nature} \textbf{\bibinfo{volume}{432}},
  \bibinfo{pages}{881} (\bibinfo{year}{2004}).

\bibitem[{\citenamefont{Dressel et~al.}(2002)\citenamefont{Dressel,
 Kasper, Petukhov, Peligrad, Gorshunov, Jourdan, Huth,
 and Adrian}}]{dre02}
\bibinfo{author}{\bibfnamefont{M.}~\bibnamefont{Dressel}},
\bibinfo{author}{\bibfnamefont{N.}~\bibnamefont{Kasper}},
\bibinfo{author}{\bibfnamefont{K.}~\bibnamefont{Petukhov}},
\bibinfo{author}{\bibfnamefont{D.~N.}~\bibnamefont{Peligrad}},
\bibinfo{author}{\bibfnamefont{B.}~\bibnamefont{Gorshunov}},
\bibinfo{author}{\bibfnamefont{M.}~\bibnamefont{Jourdan}},
\bibinfo{author}{\bibfnamefont{M.}~\bibnamefont{Huth}},
\bibnamefont{and}
\bibinfo{author}{\bibfnamefont{H.}~\bibnamefont{Adrian}},
  \bibinfo{journal}{Phys.\ Rev.\ B} \textbf{\bibinfo{volume}{66}},
  \bibinfo{pages}{035110} (\bibinfo{year}{2002}).

\bibitem[{\citenamefont{Bonn et~al.}(1988)\citenamefont{Bonn, Garrett,
 and Timusk}}]{bon88}
\bibinfo{author}{\bibfnamefont{D.~A.}~\bibnamefont{Bonn}},
\bibinfo{author}{\bibfnamefont{J.~D.}~\bibnamefont{Garrett}},
\bibnamefont{and}
\bibinfo{author}{\bibfnamefont{T.}~\bibnamefont{Timusk}},
  \bibinfo{journal}{Phys.\ Rev.\ Lett.} \textbf{\bibinfo{volume}{61}},
  \bibinfo{pages}{1305} (\bibinfo{year}{1988}).

\bibitem[{\citenamefont{Dordevic et~al.}(1988)\citenamefont{Dordevic, Basov,
 Dilley, Bauer, and Maple}}]{dor01}
\bibinfo{author}{\bibfnamefont{S.~V.}~\bibnamefont{Dordevic}},
\bibinfo{author}{\bibfnamefont{D.~N.}~\bibnamefont{Basov}},
\bibinfo{author}{\bibfnamefont{N.~R.}~\bibnamefont{Dilley}},
\bibinfo{author}{\bibfnamefont{E.~D.}~\bibnamefont{Bauer}},
\bibnamefont{and}
\bibinfo{author}{\bibfnamefont{M.~B.}~\bibnamefont{Maple}},
  \bibinfo{journal}{Phys.\ Rev.\ Lett.} \textbf{\bibinfo{volume}{86}},
  \bibinfo{pages}{684} (\bibinfo{year}{2001}).

\bibitem[{\citenamefont{Sichelschmidt et~al.}(2003)\citenamefont{Sichelschmidt,
  Ivanshin, Ferstl, Geibel, and Steglich}}]{sic03}
\bibinfo{author}{\bibfnamefont{J.}~\bibnamefont{Sichelschmidt}},
\bibinfo{author}{\bibfnamefont{V.~A.} \bibnamefont{Ivanshin}},
\bibinfo{author}{\bibfnamefont{J.}~\bibnamefont{Ferstl}},
\bibinfo{author}{\bibfnamefont{C.}~\bibnamefont{Geibel}},
\bibnamefont{and}
\bibinfo{author}{\bibfnamefont{F.}~\bibnamefont{Steglich}},
  \bibinfo{journal}{Phys.\ Rev.\ Lett.} \textbf{\bibinfo{volume}{91}},
  \bibinfo{pages}{156401} (\bibinfo{year}{2003}).

\bibitem[{\citenamefont{Ishida et~al.}(2002)\citenamefont{Ishida, Okamoto,
  Kawasaki, Kitaoka, Trovarelli, Geibel, and Steglich}}]{ish02}
\bibinfo{author}{\bibfnamefont{K.}~\bibnamefont{Ishida}},
\bibinfo{author}{\bibfnamefont{K.}~\bibnamefont{Okamoto}},
\bibinfo{author}{\bibfnamefont{Y.}~\bibnamefont{Kawasaki}},
\bibinfo{author}{\bibfnamefont{Y.}~\bibnamefont{Kitaoka}},
\bibinfo{author}{\bibfnamefont{O.}~\bibnamefont{Trovarelli}},
\bibinfo{author}{\bibfnamefont{C.}~\bibnamefont{Geibel}},
\bibnamefont{and}
\bibinfo{author}{\bibfnamefont{F.}~\bibnamefont{Steglich}},
  \bibinfo{journal}{Phys.\ Rev.\ Lett.} \textbf{\bibinfo{volume}{89}},
  \bibinfo{pages}{107202} (\bibinfo{year}{2002}).

\end{thebibliography}

\end{document}